\documentclass[reprint, amsmath,amssymb,aps]{revtex4-2}

\usepackage{graphicx}
\usepackage{dcolumn}
\usepackage{hyperref}
\usepackage{xcolor}
\usepackage{mathtools}
\usepackage{tensor}
\usepackage{siunitx}
\hypersetup{
    colorlinks,
    linkcolor={blue},
    citecolor={blue},
    urlcolor= {blue}
}
\newcommand\eftm{\mathbf{T_{{EF}}}}
\newcommand\fftm{\mathbf{C}}
\newcommand\fftma{\mathbf{C^{(1)}}}
\newcommand\fftmb{\mathbf{C^{(2)}}}
\newcommand\dtm{\mathbf{D}}
\newcommand\dop{\hat{\text{U}}_{\phi}}
\newcommand\ffeq{(52)}

\newcommand\wpf{\xi}

\newcommand\Pf{P}

\newcommand\mko{k}

\newcommand\iqb{b} 

\newcommand\lx{l_x}
\newcommand\ly{l_y} 

\newcommand\lhx[1]{x^{(#1)}} 
\newcommand\lhy[1]{y^{(#1)}}

\newcommand\cen[1]{\langle{#1}\rangle}

\DeclareMathOperator{\Tr}{Tr}

\newcommand\prs{q}

\newcommand\xg{d_x} 
\newcommand\yg{d_y}

\begin{document}

\title{Quantum Computation via Multiport Quantum Fourier Optical Processors}
\author{Mohammad Rezai}
\altaffiliation{\textit{Institute for Convergence Science and Technology} (ICST), Sharif University of Technology, Tehran 14588-89694, Iran}
\altaffiliation{\textit{Sharif Quantum Center} (SQC), Sharif University of Technology, Tehran 14588-89694, Iran}
\email{mohammad.rezai@sharif.edu}

\author{Jawad A. Salehi}
\altaffiliation{Electrical Engineering Department, Sharif University of Technology, Tehran 11155-4363, Iran}
\altaffiliation{\textit{Institute for Convergence Science and Technology} (ICST), Sharif University of Technology, Tehran 14588-89694, Iran}
\altaffiliation{\textit{Sharif Quantum Center} (SQC), Sharif University of Technology, Tehran 14588-89694, Iran}
\email{jasalehi@sharif.edu}
\date{\today}

\begin{abstract}
        The light's image is the primary source of information carrier in nature. Indeed, a single photon's image possesses a vast information capacity that can be harnessed for quantum information processing.         
         Our scheme for implementing quantum information processing via universal multiport processors employs a class of quantum Fourier optical systems composed of spatial phase modulators and 4f-processors with phase-only pupils having a characteristic periodicity that reduces the number of optical resources quadratically as compared to other conventional path encoding techniques. 
         In particular, this paper employs quantum Fourier optics to implement some key quantum logical gates that can be instrumental in optical quantum computations.
         For instance, we demonstrate the principle by implementing the single-qubit Hadamard and the two-qubit controlled-NOT gates via simulation and optimization techniques. 
         Due to various advantages of the proposed scheme, including the large information capacity of the photon wavefront, a quadratically reduced number of optical resources compared with other conventional path encoding techniques, and dynamic programmability, the proposed scheme has the potential to be an essential contribution to linear optical quantum computing and optical quantum signal processing.
\end{abstract}

\maketitle


\section{Introduction}
	Quantum light has a central role in quantum information science and technologies. 
	It culminates without substitution in various quantum information processing domains such as quantum communications~\cite{cariolaro_2015,gisin_rmp_2002} and quantum imaging~\cite{lemos_n_2014,moreau_nrp_2019}.
	Accordingly, and also due to the key role computation plays in information science and technology, optical quantum computation receives considerable attention, 
	which may eventually cause the achievement of an integrated all-optical quantum information processing system.
	These attempts have already led to various optical quantum computation approaches, such as boson sampling~\cite{aaronson_toc_2013} and linear optical quantum computing~\cite{kok_rmp_2007},
	 which rely on universal multiport unitary processors~\cite{reck_prl_1994, clements_o_2016} to process quantum optic signals.
        \par
	 Furthermore, non-photonic quantum systems such as atoms, ions, and superconducting circuits pose various challenges to quantum computation. 
	 One of the main hindrances of information processing on non-photonic quantum systems is their limited quantum life (coherence time), narrowing the allowed time for quantum information processing and computation. 
	 On the other hand, photon-based quantum computations also suffer from scalability problems due to the lack of inter-photon interaction; for example, performing controlled (entangling) gates on two single-photon qubits is challenging.
        \par
        The seminal work of Knill, Laflamme, and Milburn~\cite{knill_n_2001} (KLM) proposed a solution to this problem of quantum computation by photons.
        It introduces universal quantum computing based on single-photon sources, number-resolving photodetectors, and linear optical elements.
        The operation on the quantum state of single photons is implemented via linear optical elements, namely, beam-splitters and phase-shifters.
        In the KLM protocol, projective measurement at the output of the linear optical operation, probabilistically, enables and also heralds the entangling gates.
        In addition, it presents a near-deterministic teleportation scheme with linear optics, enhanced by quantum error correction coding, to achieve scalable teleportation-based quantum computing~\cite{gottesman_n_1999}.
        \par
        Despite the high challenges and resource demands of KLM protocols, it gives a good insight into how to achieve optical universal quantum computation.
       The KLM protocols and various linear optical quantum information processing schemes rely heavily on the fact that any $N\times N$~unitary operation on the multi-optical-ports is realizable via a sequence of, at most,~$\frac{N(N-1)}{2}$ beam-splitters and phase-shifters~\cite{reck_prl_1994}.
       Due to the interferometric nature of such linear optical unitary processors, the increase in the optical depth and number of optical devices for implementation can significantly reduce the fidelity and success of quantum operations in practice. 
       Therefore, several attempts have been made to reduce the optical depth and required number of optical elements~\cite{clements_o_2016, lukens_o_2017}.
        \par
        	Matrix factorization analysis~\cite{muellerquade_phd_1998,huhtanen_jfa_2015} shows that any $N\times N$~matrix is the product of a sequence of at most $2N-1$ alternating diagonal and circulant matrices, quadratically fewer than the corresponding matrix factorization into the product of beam-splitter and phase-shifter matrices, which is of order $\mathcal{O}(N^2)$.
        Since the discrete Fourier transform~$F$ can diagonalize circulant matrices~\cite{robert_cit_2006}, an alternating circulant~($C$) and diagonal~($D$) matrices product is equivalent to the alternating discrete Fourier transform and diagonal matrices product, more precisely, $C_1D_1\hdots D_{N-1} C_{N}=FD'_1F^{\dagger}D_1F\hdots   D_{N-1} FD'_{N}F^{\dagger}$, where $D'_j$ is the diagonal eigenvalue matrix of $C_j$, i.e.,~$C_j=FD'_jF^{\dagger}$.  
         In other words, alternating diagonal operation on Fourier dual spaces such as time and frequency~\cite{lukens_o_2017} can render any unitary transformation.
         \par
         Fourier optics is an alternative approach to implementing any unitary transformation via an alternating diagonal operation on the Fourier dual spaces~\cite{muellerquade_phd_1998,morizur_osa_2010,zhao_joo_2019,lopez_oe_2021}.
         Accordingly, quantum Fourier optics~\cite{qfo} provides a platform to implement quantum information processing and quantum computing.  
          Spatial light modulation (SLM) technology allows Fourier optical quantum computation to be electronically programmable~\cite{jacques_s_2015,bogaerts_n_2020,chi_nc_2022}, a feature that offers many promises. 
         Moreover, due to the two-dimensional wavefront space, as opposed to the one-dimensional frequency comb techniques limited to the operating frequency range of the involved optical elements, Fourier optical quantum information processing is more desirable for scalability purposes.
         More importantly, programmable Fourier optical quantum information processing, compared to the Mach-Zehnder-interferometer-based path encoding with beam-splitters and phase shifters~\cite{jacques_s_2015}, requires quadratically fewer devices~\cite{reck_prl_1994,huhtanen_jfa_2015} and is more stable over time~\cite{li_prad_2020}, enhancing the practical fidelity of the implemented quantum gates. 
         \par
         Recent work on the fundamentals of quantum Fourier optics~\cite{qfo} has developed the required mathematical models and tools for Fourier optical quantum computation. 
         This paper employs these mathematical models to introduce a novel quantum information processing based on the photon’s wavefront information capacity.
          Initially, we will study quantum Fourier optics, focusing on discretized photon wavefronts. 
          Quantum information processing on discretized wavefronts is practically feasible using lattice-like optical fiber arrays and specific periodic-only spatial phase modulation, making quantum Fourier optics an attractive option for discrete variable linear optical quantum computation~\cite{kok_rmp_2007}. 
          In particular, we show that 4f-processors with a generic class of periodic pupil phase factors preserve the space discretization at its output. 
          Furthermore, the matrix representation of such a 4f-optical configuration corresponds to the circulant and diagonal matrices product. 
          This factorization of the 4f-operator is the building block to making any arbitrary unitary transformation. 
          Finally, we implement a set of Fourier optical-based quantum gates, namely the single-qubit Hadamard gate and the two-qubits entangling C-Not gate. 
        \section{Quantum Fourier Optics}
        Classical Fourier optics studies the light’s wavefront transformation while propagating through optical systems composed of lenses and spatial light modulators and filters. 
        In addition to the photon wavefront evolution, quantum Fourier optics also includes the Fock representation of the quantum light. 
        Indeed, when the input of the optical system is coherent (Glaber) states or the input light is composed of photons with identical wavepackets (wavefronts), classical and quantum Fourier optics yield the same result. 
         \par
        The fundamentals of quantum Fourier optics~\cite{qfo} demonstrate the evolution of a generic class of pure quantum states~$\lvert \psi \rangle  =f(\hat a_{\wpf}^ \dagger) \lvert 0 \rangle$~\cite{glauber, rezai_ieeeit_2021}
        through Fourier optical systems, where function $f$ is an arbitrary analytic (infinitely differentiable), normalized function of the photon-wavepacket creation operator~$\hat a_{\wpf}^ \dagger$ ($\wpf$ denotes the shape of the photon-wavepacket)
        \begin{equation}
          \begin{split}
            \lvert \psi \rangle &=f(\hat a_{\wpf}^ \dagger) \lvert 0 \rangle =\sum_{n=0}^{\infty} c_n \frac{\hat a_{\wpf}^{\dagger n}}{\sqrt{n!}}\lvert 0 \rangle= \sum_{n=0}^{\infty}c_n \lvert n \rangle_{\wpf}\, ,
          \end{split}
          \label{eqn:si=f(adag)}
        \end{equation}
        where~$c_n$ corresponds to the~$n$th Taylor coefficient of function~$f(\hat a_{\wpf}^ \dagger)$, $\lvert 0 \rangle$ is the vacuum state, and~$\lvert n \rangle_{\wpf}$ represents the $n$-photon Fock state with wavepacket~$\wpf$.
        \par
        This paper assumes the input photons of the Fourier optical systems occupy an identical single spectral mode with angular frequency~$\omega$ and an identical polarization mode with polarization~$p$.
        Consequently, under paraxial approximation, the only degree of freedom for photons' occupation mode is their normalized wavefront, shown by symbol~$\wpf$
        \begin{equation}
            \iint dx dy   \,  \lvert \wpf (x,y) \rvert^2 =1\, .
          \label{eqn:wpnorm}
        \end{equation}
  	Therefore, the photon-wavepacket creation operators~\cite{loudon_2000} take the photon-wavefront as their subscript.
        In other words, the creation operator~$\hat a_{\wpf}^ \dagger$ creates a single photon with wavefront~$\wpf (x,y)$
        \begin{equation}
            \hat a^ \dagger _{\wpf} =\iint dx dy   \,  \wpf (x,y) \hat a^\dagger_{x,y} \, ,
          \label{eqn:axixy}
        \end{equation}
        where $\hat a^\dagger_{x,y}$ creates a photon at the position with Cartesian coordinates~$x$ and $y$.
        It obeys the following canonical commutation relation        
         \begin{equation}
           \left[\hat a_{x,y},\hat a^\dagger_{x',y'}\right] =\delta(x-x')\delta(y-y') \, .
           \label{eqn:[a,ad]}
        \end{equation}
        \par
        Note that the quantum state of a single-photon, $f( \hat a^ \dagger _{\wpf} )= \hat a^ \dagger _{\wpf} $, according to Eq.~\eqref{eqn:si=f(adag)} and~\eqref{eqn:axixy}, is as follows
         \begin{equation}
          \begin{split}
            \lvert \psi \rangle &=   \hat a^ \dagger _{\wpf}    \lvert 0 \rangle 
            =\iint dx dy   \,  \wpf (x,y)   \lvert 1 \rangle_{x,y} \, .
            \end{split}
          \label{eqn:}
        \end{equation}
 
        \subsection{Discrete Spatial Modes}
        \begin{figure}[!t]
          \centering
          \includegraphics[width=\columnwidth]{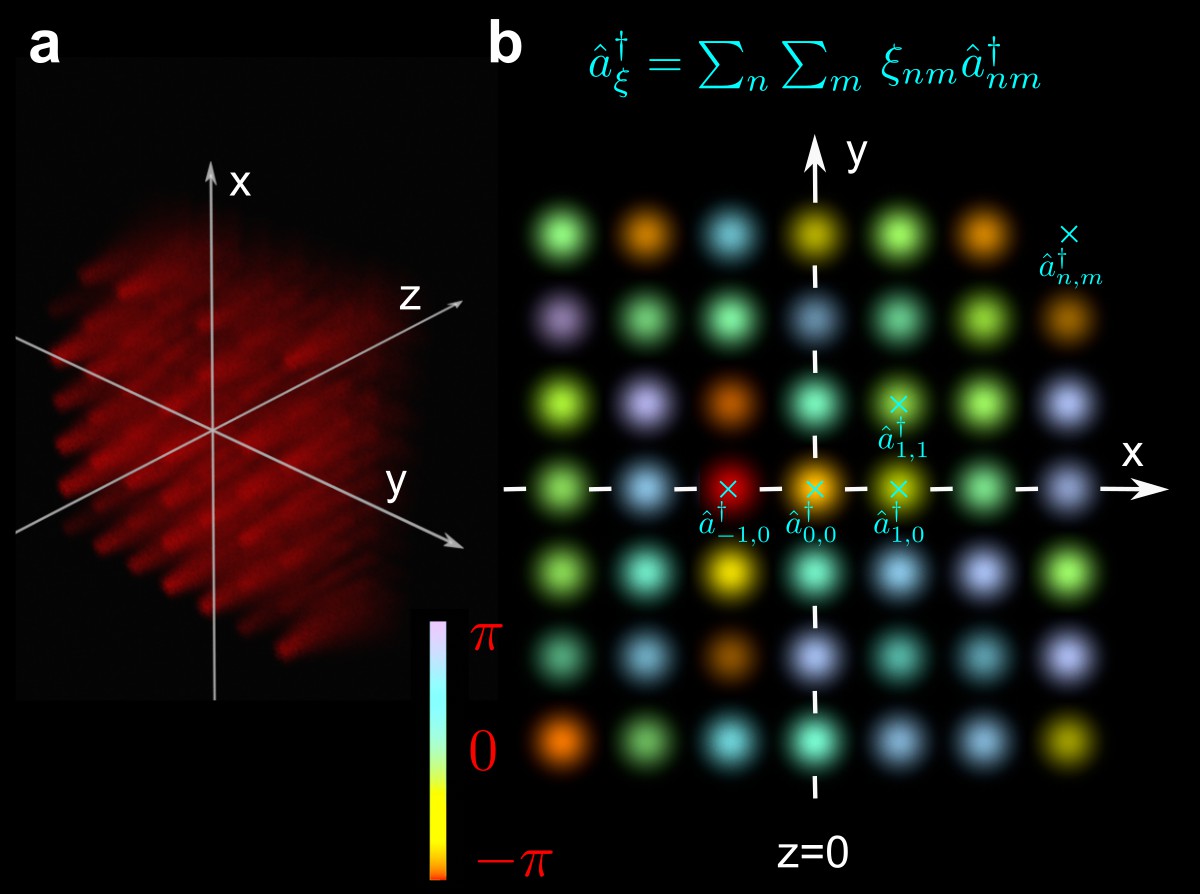}
          \caption{\textbf{Discretized (Digitalized) Photon-Wavepacket.}
          \textbf{a} shows the propagation of a single photon in the~$z$-direction, which has a lattice-like wavefront at $z=0$.
          \textbf{b} shows the photon's wavefront of~\textbf{a} at $z=0$.
          Its phase is color-coated according to the color map in the lower-left corner.
          The intensity of the colors corresponds to the amplitude of the photon-wavefront.
          }\label{fig:diswp}
        \end{figure}
        In this paper, we consider the photon wavefront as the carrier of quantum information.
        To encode quantum information into the photon wavefronts as a discrete-mode (discrete-variable) such as qubits or generally qudits, we assume at $z=0$ the photon-wavepacket~${\wpf}$ can have a nonzero amplitude only at the lattice-points separated by $\lx$ and $\ly$ in the $x-y$~plane (see Fig.~\ref{fig:diswp}).
        Let $  \wpf_{nm}$ denote the single-photon probability amplitude at the lattice-point~$x=n \lx,y=m \ly$. 
        Therefore, similar to Eq.~\eqref{eqn:wpnorm}, the normalization condition of a single-photon quantum state implies that 
        \begin{equation}
          \begin{split}
            \sum_n\sum_m  \lvert \wpf_{nm}\rvert^2=1\,.
          \end{split}
          \label{eqn:diswpnorm}
        \end{equation}
        The creation operator for such a lattice-like photon-wavepacket is representable as
        \begin{equation}
          \begin{split}
            \hat a^ \dagger _{\wpf}            &=\sum_n\sum_m    \,  \wpf_{nm} \hat a^\dagger_{nm}\, ,
          \end{split}
          \label{eqn:axinm}
        \end{equation}
        where $\hat a^\dagger_{nm}$ denotes the photon creation operator in position $x=n \lx,y=m \ly$, and their commutation relations are as follows:
         \begin{equation}
          \begin{split}
            \left[\hat a_{nm},\hat a^\dagger_{n'm'}\right]& =\delta_{nn'}\delta_{mm'}\, ,
          \end{split}
          \label{eqn:[a,ad]nm}
        \end{equation}
        where $\delta$ denotes the Kronecker delta function.
        It is worth noting that equations~\eqref{eqn:diswpnorm} to~\eqref{eqn:[a,ad]nm} of the discrete spatial modes correspond to Eq~\eqref{eqn:wpnorm} to~\eqref{eqn:[a,ad]} of the continuous spatial mode, respectively.
        Equation~\eqref{eqn:axinm} adopts the \textit{square} lattice-like wavefront only for the sake of simplicity.
        The formalism is easily extendable to other two-dimensional lattices, such as rectangular and hexagonal lattices. 
        \par
        The quantum state of a single-photon, $f( \hat a^ \dagger _{\wpf} )= \hat a^ \dagger _{\wpf} $, is, according to Eq.~\eqref{eqn:si=f(adag)} and~\eqref{eqn:axinm}, expressible as
        \begin{equation}
          \begin{split}
            \lvert \psi \rangle &=   \hat a^ \dagger _{\wpf}    \lvert 0 \rangle 
            =\sum_n\sum_m    \,  \wpf_{nm}  \lvert 1 \rangle_{nm} \, .
            \end{split}
          \label{eqn:qudit}
        \end{equation}
        The equation above indicates $d$-dimensional quantum information encoded on a single-photon, a qudit.
        Let us consider $N$ lattice points to the $x$-direction and $M$ lattice points to the $y$-direction for photon occupation.
        Therefore, the $d$-dimension of a single-photon Hilbert space equals~$d=NM$, for example in Fig~\ref{fig:diswp}, $N=M=7$, and $d=49$.
        \par        
        Practically, the introduced discrete spatial point creation operator~$\hat a^\dagger_{nm}$ is associated with a quantum optical source with a central point at $(\langle x \rangle ,\langle y \rangle) = ( n\lx  , m \ly )$ and its diffraction-limited waist confined in the area of size~$\lx \ly $.
        So, there is no overlap between the sources associated with different lattice points, and therefore the commutation relation~\eqref{eqn:[a,ad]nm} holds.
        \par
        Optical fiber arrays and quantum-emitter arrays~\cite{berraquero_nc_2017} can be utilized to realize lattice-like wavefronts,  making the proposed scheme practical and interesting for various discrete-modes (discrete-variable) quantum information encoding techniques.
An effective quantum Fourier operation for such a discretized wavefront is a class of 4f-processors preserving the discretization of the wavefront at their output. The following section presents details about this topic.
      \section{Circulant Matrix Operation via 4\lowercase{f}-Processing System}
      \begin{figure}[!t]
          \centering
          \includegraphics[width=\columnwidth]{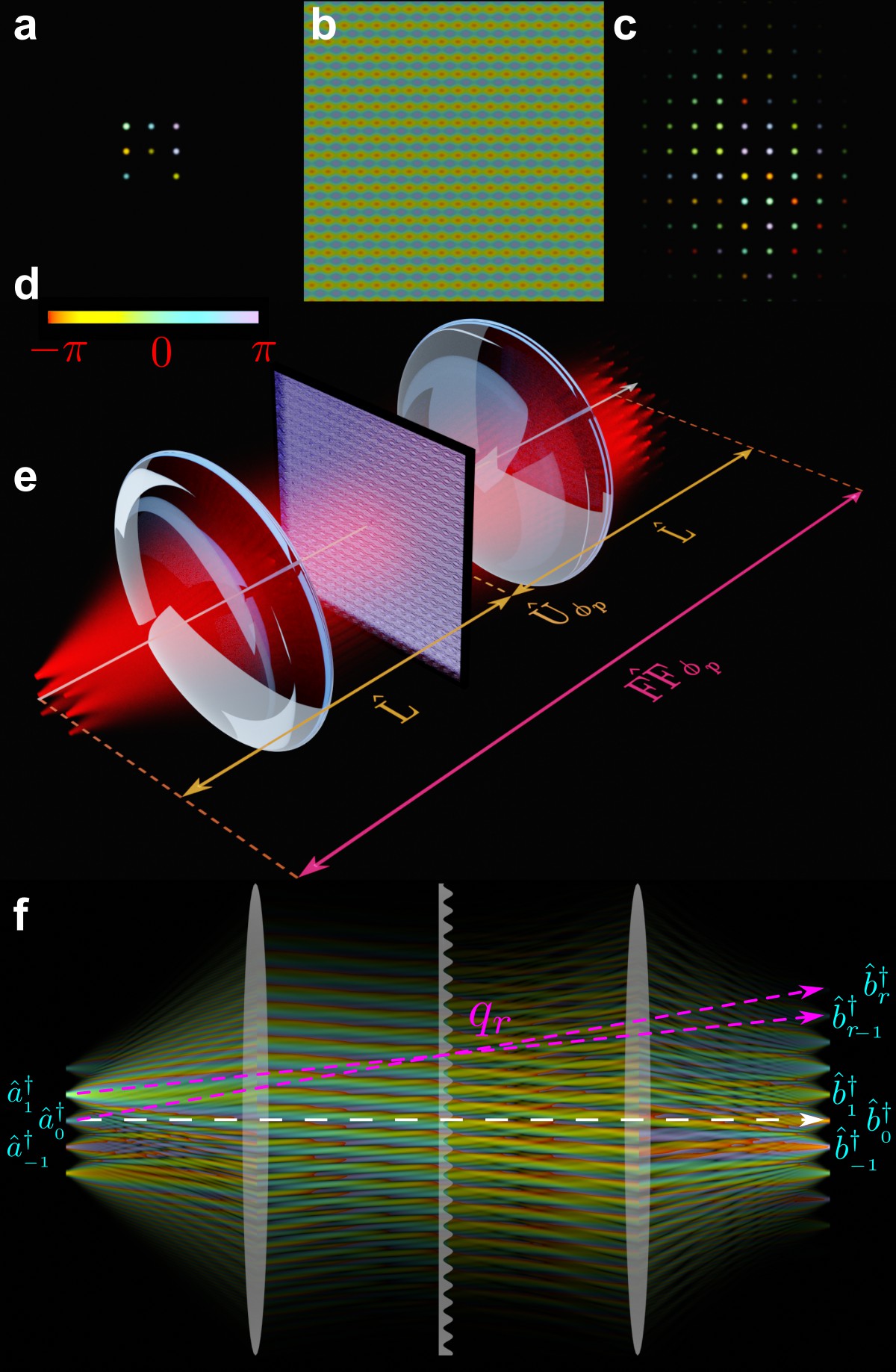}
          \caption{\textbf{Circulant Matrix Operation via 4f-Processing Systems.}
          \textbf{e} shows a 4f-processing system composed of two lens operators and one spatial phase modulator (see~\cite{qfo} for its quantum modeling details) and simulates the intensity propagation of a single photon through the system.
          \textbf{a} displays the discretized photon-wavefront at the input. 
          The phase modulation of the 4f-system's pupil can be seen in \textbf{b}.
          Since it has a  specific spatial periodicity, the photon-wavefront at the output (\textbf{c}) has a lattice structure similar to the input.
          In this figure, colormap~\textbf{d} is used to show the phases of the modulator and photon wavefront.
          Color intensity corresponds to wavefront amplitude.
          \textbf{f} shows a one-dimensional photon-wavefront propagating through a 4f-processor.
          It illustrates the transformation of creation field operators associated with the pupil phase modulator's Fourier expansion coefficientt~$q_r$.
          For each point~$z$ of the propagation, we show only the relative phases and ignore the no-observable net displacement phase factor~$e^{ikz}$. }\label{fig:4fdis}
        \end{figure}
        Consider a 4f-processing system that comprises two identical lenses with focal length~$f$ and a pupil on the confocal plane (see Fig.~\ref{fig:4fdis}).
        Assume the pupil is a periodic phase-only spatial modulator~$\Pf(x,y)= e^{-i \phi_p(x,y)}$ with periods~$\xg$ and $\yg$ (spatial angular frequencies $\kappa_x=\frac{2 \pi }{\xg}$ and $\kappa_y=\frac{2 \pi }{\yg}$) in the $x$ and $y$ directions, respectively.
        Therefore, the pupil's Fourier series expansion is as follows
         \begin{equation}
          \begin{split}
            \Pf(x,y)= e^{-i \phi_p(x,y)} &=\sum_r\sum_s \prs_{rs} e^{i\left( r\kappa_x x + s\kappa_y y\right)} \, ,\\
          \end{split}
          \label{eqn:ft[ffp]main}
        \end{equation}
        where $\prs_{rs}$ is the two-dimensional Fourier expansion's coefficient of the periodic pupil phase factor.
        \par
        As shown in ~\cite{qfo} (Eq.~\ffeq), the corresponding quantum operator~$ \hat{\text{FF}}_{\phi_p}$ of such a 4f-system transforms creation operator $\hat a^ \dagger _{\cen{x},\cen{y}}$ associated with a localized source at coordinate~$(\cen{x},\cen{y})$ on the input plane is as follows 
        \begin{equation}
          \begin{split}
            \hat{\text{FF}}_{\phi_p}\, \hat a^ \dagger _{\cen{x},\cen{y}}\, \hat{\text{FF}}^{\dagger}_{\phi_p}
            &= 
           \sum_r\sum_s  \prs_{rs} \,  \hat b^\dagger_{ \lhx{r}-\cen{x}, \lhy{s} - \cen{y}}\, ,
            \end{split}
          \label{eqn:ffadxyffperiodic}
        \end{equation}
        where
        \begin{equation}
          \begin{split}
           \lhx{r} &=r\lhx{1}=r\frac{f \kappa_x}{\mko}\\
          \lhy{s}&=s\lhy{1}=s\frac{f \kappa_y}{\mko} \, ,
        \end{split}
        \label{eqn:lxy'}
      \end{equation}
      $f$ is the lenses' focal length, and $k$ is the photon wavevector.
      In Eq.~\eqref{eqn:ffadxyffperiodic}, we have dropped the trivial constant phase factor~$-e^{i 4 \mko f}$ (a global phase factor), which does not affect the overall shape of the photon wavepacket and the corresponding quantum state.
        This global phase shift is due to the net displacement through the 4f-system.
        However, to be precise, one may pull the phase by adding it to the transformation coefficient~$\prs_{rs}$.
      Equation~\eqref{eqn:ffadxyffperiodic} denotes that the output of such a 4f-processor is lattice-like with lattice constants~$\lhx{1}=\frac{f \kappa_x}{\mko}$ and $\lhy{1}=\frac{f \kappa_y}{\mko}$ in the $x$ and $y$ directions, respectively.
      Therefore, to extend and adapt Eq.~\eqref{eqn:ffadxyffperiodic} for lattice-like input quantum states with photon creation operator Eq.~\eqref{eqn:axinm} and match the output lattice with the input lattice, we assume the pupil's spatial angular frequencies for the $x$ and $y$ directions are $\kappa_x=\frac{2 \pi }{\xg}= \frac{\mko \lx}{f}$ and $\kappa_y=\frac{2 \pi }{\yg}= \frac{\mko \ly}{f}$, respectively.
      This choice of periods for the pupil makes the 4f output lattice-constants equal the input lattice-constants, i.e., $\lhx{1}=\lx$ and $\lhy{1}=\ly$.
      Accordingly, we use the same formalism as Eq.~\eqref{eqn:axinm} for the output creation operators and write $ \hat b^\dagger_{ \lhx{r}-\cen{x}, \lhy{s} - \cen{y}}$  for $\cen{x}= n \lx$ and $\cen{y}=m\ly$ as  $\hat b^\dagger_{r-n, s-m}$.
      Thus, Eq.~\eqref{eqn:ffadxyffperiodic} becomes
      
        \begin{equation}
          \begin{split}
            \hat{\text{FF}}_{\phi_p} \hat a^\dagger_{nm} \hat{\text{FF}}^{\dagger}_{\phi_p} = &\sum_r\sum_s \prs_{rs} \hat b^\dagger_{r-n,s-m}\\
            =&\sum_r\sum_s \prs_{n+r,m+s} \hat b^\dagger_{r,s}\, ,
            \end{split}
          \label{eqn:4fop}
        \end{equation}
        where, in the second line, $r$ is substituted with $n+r$ and $s$ with $m+s$.
        \par
        Equation~\eqref{eqn:4fop} (see also Fig.~\ref{fig:4fdis}) indicates that the 4f-processor with a periodic phase-only pupil transforms each input lattice-point creation operator~$\hat a^\dagger_{nm}$ into a linear combination of output lattice-point creation operators~$\hat b^\dagger_{r,s}$.
        The corresponding transformation amplitude is given by $\prs_{n+r,m+s}$, which are the Fourier coefficients of the pupil phase factor of the 4f-processor (Eq.~\eqref{eqn:ft[ffp]main}).
        \subsection{One Dimensional Quantum 4f-Transformation}
        For the sake of simplicity, in the rest of the paper, we consider one-dimensional wavefronts for the photon-wavepacket.
        This assumption reduces Eq.~\eqref{eqn:4fop} to 
        \begin{equation}          
            \hat{\text{FF}}_{\phi_p} \hat a^\dagger_{n} 
            \hat{\text{FF}}^{\dagger}_{\phi_p}=  \sum_r\prs_{r} \hat b^\dagger_{r-n}=\sum_r \prs_{n+r} \hat b^\dagger_{r}\, .
          \label{eqn:4fop1d}
        \end{equation}
        The above transformation has the following matrix representation:
        \small
        \begin{subequations}
          \begin{gather}
           \label{eqn:heispicffa}
            \hat{\text{FF}}_{\phi_p}
            \begin{pmatrix}
              \vdots\\
              \hat a^{\dagger}_{2}\\
              \hat a^{\dagger}_{1}\\
              \hat a^{\dagger}_{0}\\
              \hat a^{\dagger}_{-1}\\
              \hat a^{\dagger}_{-2}\\
              \vdots 
            \end{pmatrix}
             \hat{\text{FF}}^{\dagger}_{\phi_p}
             =            
            \begin{pmatrix}
              \ddots &\vdots & \vdots & \vdots &\vdots &\vdots &\reflectbox{$\ddots$}\\
              \hdots & \prs_{4}  & \prs_{3}  & \prs_{2}   & \prs_{1}  &\prs_{0}  &\hdots\\
              \hdots & \prs_{3}  & \prs_{2}  & \prs_{1}   & \prs_{0}  &\prs_{-1}  &\hdots\\
              \hdots & \prs_{2}  & \prs_{1}  & \prs_{0}   & \prs_{-1}  &\prs_{-2}  &\hdots\\
              \hdots & \prs_{1}  & \prs_{0}  & \prs_{1}   & \prs_{-2}  &\prs_{-3}  &\hdots\\
              \hdots & \prs_{0}  & \prs_{-1}  & \prs_{-2}   & \prs_{-3}  &\prs_{-4}  &\hdots\\
              \reflectbox{$\ddots$} &\vdots & \vdots & \vdots &\vdots &\vdots &\ddots             
            \end{pmatrix}\,
            \begin{pmatrix}
               \vdots\\
              \hat b^{\dagger}_{2}\\
              \hat b^{\dagger}_{1}\\
              \hat b^{\dagger}_{0}\\
              \hat b^{\dagger}_{-1}\\
              \hat b^{\dagger}_{-2}\\
              \vdots 
            \end{pmatrix} \, ,\\
        \intertext{and in the short form, it is}   
        \label{eqn:heispicffb}
        \hat{\text{FF}}\mathbf{\hat a^{\dagger}} \hat{\text{FF}}_{\phi_p}=\fftm \mathbf{.\hat b^{\dagger}}\, ,          
          \end{gather}
          \label{eqn:heispicff}
        \end{subequations}
        \normalsize
        where $\mathbf{\hat a^{\dagger}}$ and $\mathbf{\hat b^{\dagger}}$ are column vectors with elements of creation operators $ \hat a^\dagger_{n}$ and $\hat b^\dagger_{r}$, respectively, and $\fftm $, the 4f-transformation matrix~$\fftm $ is a circulant-type matrix with elements 
        \begin{equation}
          (\fftm )_{n,r}=\prs_{n+r}\, .
        \end{equation}
      Furthermore, as is shown in~\cite{qfo}, the pupil phase factor's Fourier  coefficients~$\prs_r$ are cyclic orthogonal, and therefore, $\fftm$ is a unitary circulant matrix.
      \par
        To conclude, the 4f-transformation with the appropriate periodic pupil phase factor keeps the discreteness and the lattice-like structure of the input wavefront.
        Therefore, matrix multiplication can represent the 4f-transformation. 
        Furthermore, the 4f-transformation matrix is a unitary circulant matrix.

\section{Universal Multiport Operation}           
        As mentioned earlier, any $N\times N$~matrix~$\mathbf{T}$ is factorizable to a product of at most $2N-1$ alternating  $N$~circulant and $N-1$ diagonal matrices~\cite{huhtanen_jfa_2015}, i.e., $\mathbf{T}=\mathbf{C^{(1)}.D^{(1)}\hdots D^{(N-1)}.C^{(N)}}$, hence implementable via a Fourier optical system~\cite{muellerquade_phd_1998}.
        \par
        The previous section (Eq.~\eqref{eqn:heispicff}) shows that a one-dimensional 4f-system quantum operator with the periodic pupil performs a circulant matrix operation on the input single-photons' lattice-like encoded quantum information.
        Furthermore, a spatial phase modulation operator performs the diagonal matrix operation~\cite{qfo}.
        The quantum operator of a lattice-like spatial phase modulator is as follows
        \begin{equation}
          \dop  = e^{- i \sum_{r,s} \phi_{rs} \hat n_{rs}} =\prod_{r,s} e^{- i \phi_{rs} \hat n_{rs}}\, ,
          \label{eqn:Uphinm}
        \end{equation}
        where $\hat n_{rs}=\hat b^{\dagger}_{rs}\hat b_{rs}$ is the number operator associated with the quantum state of the two-dimensional lattice point~$(r,s)$.
        The phase~$\phi_{rs}$ is the phase the spatial phase modulator applies to the lattice cell with a central point at coordinate~$(r\lx,s\ly)$.  
        One can drop the subscript~$s$ in Eq.~\eqref{eqn:Uphinm} for a one-dimensional lattice, transforming the one-dimensional lattice creation operator~$\hat b^{\dagger}_{r}$ as follows
        \begin{equation}
          \begin{split}
            \dop \hat b^{\dagger}_{r} \dop ^{\dagger}  &=\left(\prod_{r'} e^{ -i \phi_{r'} \hat n_{r'}} \right) \hat b^{\dagger}_{r}\left(\prod_{r''} e^{ i \phi_{r''} \hat n_{r''}}\right)\\
            &= e^{ -i \phi_{r} \hat n_{r}}  \hat b^{\dagger}_{r} e^{ i \phi_{r} \hat n_{r}} \\
            &=   e^{ -i \phi_{r} } \hat b^{\dagger}_{r} \\
            &=   \dtm_{rr} \hat b^{\dagger}_{r} \, ,
            \end{split}
          \label{eqn:ubud}
        \end{equation}
        where~$\dtm_{rr}=e^{ -i \phi_{r} }$ is the $r$th diagonal element of the diagonal transformation matrix~$\dtm$ associated with the one-dimensional spatial phase modulation operator~$ \dop$.
        \par
        To sum up, Fourier optics via a cascade of 4f-operators~$\hat{\text{FF}}$,  lattice-like spatial modulation operators~$\dop $, whose transformation matrices are circulant matrices and diagonal matrices, respectively, can realize universal multiport transformations.
        In the following section, based on this factorization principle, we demonstrate quantum gates in the quantum Fourier optical platform using an 8f-processing system, a cascade of two 4f-processors interconnected with a lattice-like spatial modulator.
        Therefore, let us consider an 8f-processor in more detail.
        \subsection{Quantum Based 8\lowercase{f}--Processor}
        In the 8f-processor, the output of the first 4f-processor~$\hat{\textbf{FF}}^{(1)}$ (Eq.~\eqref{eqn:4fop1d}) goes to a lattice-like spatial phase modulator~$\dop $ (Eq.~\eqref{eqn:ubud}), which gives the following transformation
        \begin{equation}
          \begin{split}
          \dop  \left(\hat{\text{FF}}^{(1)} \hat a^\dagger_{n} 
            \hat{\text{FF}}^{(1) \dagger} \right)\dop ^{\dagger}&= \sum_r\prs^{(1)}_{n+r} \dop  \hat b^\dagger_{r} \dop ^{\dagger}\\
          &= \sum_r\prs^{(1)}_{n+r} e^{ -i \phi_{r} } \hat b^\dagger_{r} \, .
          \end{split}
          \label{eqn:u4fop1d}
        \end{equation}
        The superscript~$(1)$ is added to the transformation coefficient~$\prs^{(1)}_{n+r}$ of the first 4f-operator~$\hat{\text{FF}}^{(1)}$ to discriminate them from the transformation coefficient~$\prs^{(2)}$ of the second 4f-operator~$\hat{\text{FF}}^{(2)}$.
        Using Eq.~\eqref{eqn:4fop1d}, the transformation by operator~$\hat{\text{FF}}^{(2)}$ takes  the following form
        \begin{equation}          
            \hat{\text{FF}}^{(2)} \hat b^\dagger_{r} 
            \hat{\text{FF}}^{(2)\dagger}
            =\sum_l \prs^{(2)}_{r+l} \hat d^\dagger_{l}\, .
          \label{eqn:4f2op1d}
        \end{equation}
        Considering Eq.~\eqref{eqn:u4fop1d} and~\eqref{eqn:4f2op1d}, the 8f-processing system operator
        \begin{equation}
        \hat{\text{EF}}= \hat{\text{FF}}^{(2)} \dop  \hat{\text{FF}}^{(1)}
        \end{equation}
        is the cascade of 4f-operator~$\hat{\text{FF}}^{(1)}$, spatial modulation operator~$\dop $, and 4f-operator~$ \hat{\text{FF}}^{(2)}$, which performs the following transformation
        \begin{equation}
          \begin{split}
             \hat{\text{EF}} \hat a^\dagger_{n} \hat{\text{EF}}^{\dagger}&=
          \hat{\text{FF}}^{(2)} \left( \dop\,  \hat{\text{FF}}^{(1)} \hat a^\dagger_{n} 
            \hat{\text{FF}}^{(1) \dagger}\, \dop ^{\dagger}\right) \hat{\text{FF}}^{(2)\dagger}\\
          &= \sum_r\prs^{(1)}_{n+r} e^{ -i \phi_{r} } \hat{\text{FF}}^{(2)} \hat b^\dagger_{r}  \hat{\text{FF}}^{(2)\dagger}\\
          &= \sum_{r,l}\prs^{(1)}_{n+r} e^{ -i \phi_{r} } \prs^{(2)}_{r+l}   \hat d^\dagger_{l}\\
          &= \sum_{r,l}\fftma _{nr}  \dtm_{rr} \fftmb _{rl}   \hat d^\dagger_{l}\, ,
          \end{split}
          \label{eqn:efanef}
        \end{equation}
        where $\fftma _{nr}=\prs^{(1)}_{n+r} $ and $\fftmb _{rl}=\prs^{(2)}_{r+l}$ are the unitary circulant transformation matrix elements of the 4f-operators~$\hat{\text{FF}}^{(1)}$ and $\hat{\text{FF}}^{(2)}$, respectively.
        Also, $\dtm_{rr'}= e^{ -i \phi_{r} } \delta_{rr'}$ is the $r$th element of the unitary diagonal transformation matrix of the spatial phase modulation operator~$\dop ^{\dagger}$.
        Therefore, the matrix representation of the above 8f-processor is as follows
        \begin{equation}
          \begin{split}
            \hat{\text{EF}}\mathbf{\hat a^{\dagger}} \hat{\text{EF}}^{\dagger}&=\mathbf{ \eftm. \hat d^{\dagger}}\, ,
          \end{split}
          \label{eqn:8fprocessor}
        \end{equation}
        where
        \begin{equation}
          \eftm=\mathbf{\fftma . \dtm. \fftmb }
          \label{eqn:8ftm}
        \end{equation}
        is the transformation matrix of the 8f-operator~$\hat{\text{EF}}$, and $\mathbf{\hat a^{\dagger}}$ and $\mathbf{\hat d^{\dagger}}$ are column vectors with elements of input creation operators~$\hat a^\dagger_{n}$ and output creation operator~$\hat d^\dagger_{l}$ of the 8f-processor.
        \section{Quantum Computation via Quantum Fourier Optics}
        The previous section demonstrates the power of Fourier optical systems in implementing universal multiport (discrete-variable) unitary operations~\cite{reck_prl_1994,clements_o_2016,jacques_s_2015}.
        Accordingly, quantum Fourier optical systems offer a platform for linear optical quantum computations such as KLM scheme~\cite{knill_n_2001} and boson sampling~\cite{aaronson_toc_2013}.
        It is shown that a product of two circulant matrices and a diagonal matrix can produce universal quantum gates~\cite{lukens_o_2017,kues_np_2019}, which corresponds to an 8f-processor transformation matrix (Eq.~\eqref{eqn:8ftm}).
        Accordingly, this section uses an 8f-processor to implement the single-qubit Hadamard gate and the two-qubit controlled-NOT (CNOT) entangling gate, two crucial gates for quantum computation~\cite{nielsen_chuang_2010}.
        \par
        In our Fourier optical formalism, a qubit is a photon in a superposition of two different lattice points~$m$ and~$m'$.
        This paper considers single-photon occupation in the neighboring points~$m$ and $m'=m+1$ as the two \textit{computational basis states} of the qubit.
        We show these basis states as~$\lvert \Downarrow \rangle=\lvert 1 \rangle_{m}=\hat a^{\dagger}_m\lvert 0\rangle$ and~$\lvert \Uparrow \rangle=\lvert 1 \rangle_{m+1}=\hat a^{\dagger}_{m+1}\lvert 0\rangle$, respectively.
        Therefore, in this formalism, the quantum state of the  ${\iqb}$th qubit is expressible by a discretized photon-wavefront creation operator Eq.~\eqref{eqn:axinm} as follows
        \begin{equation}
          \begin{split}
            \lvert \psi^{({\iqb})} \rangle &= \xi^{({\iqb})}_{\Downarrow} \lvert \Downarrow \rangle_{{\iqb}} +\xi^{({\iqb})}_{\Uparrow} \lvert \Uparrow \rangle_{{\iqb}} \\
            &= \xi_{m_{\iqb}} \lvert 1 \rangle_{m_{\iqb}}+ \xi_{m_{\iqb}+1} \lvert 1 \rangle_{m_{\iqb}+1}\\
            &=\hat a^{ ({\iqb})\dagger}_{\xi} \lvert 0 \rangle\, .
          \end{split}
          \label{eqn:psib}
          \end{equation}
          Assume the Fourier optical system is composed of $B$~qubits.
          Since the lattice points of each qubit should be different from other qubits, we consider the two states of the ${\iqb}$th qubit as~$ \lvert \Downarrow \rangle_{{\iqb}}=\lvert 1 \rangle_{2{\iqb}}$ and ~$\lvert \Uparrow \rangle_{{\iqb}}=\lvert 1 \rangle_{2{\iqb}+1}$, which means~$m_{\iqb} =2 {\iqb}$, in Eq.~\eqref{eqn:psib}.
          Therefore, the state of $B$ separable (non-entangled) qubits takes the following form
          \begin{equation}
          \begin{split}
            \lvert \Psi \rangle &= \prod_{{\iqb}}\lvert \psi^{({\iqb})} \rangle \\
            &=\prod_{{\iqb}}\Big(\xi^{({\iqb})}_{\Downarrow} \lvert \Downarrow \rangle_{{\iqb}} +\xi^{({\iqb})}_{\Uparrow} \lvert \Uparrow \rangle_{{\iqb}}\Big)\\
            &= \prod_{{\iqb}}\Big(\xi_{2{\iqb}} a^{\dagger}_{2{\iqb}} +  \xi_{2{\iqb}+1}a^{\dagger}_{2{\iqb}+1}  \Big)\lvert 0 \rangle\, .
          \end{split}
          \label{eqn:ef_in}
          \end{equation}
        If the quantum state Eq.~\eqref{eqn:ef_in} inters 8f-processor~$\hat{\text{EF}}$ with transformation matrix~$\eftm$, the output state would be
        \begin{equation}
          \begin{split}
            \lvert \Phi \rangle &= \hat{\text{EF}} \lvert \Psi \rangle\\
            &=\hat{\text{EF}} \prod_{{\iqb}}\lvert \psi^{({\iqb})} \rangle\\
            &=\hat{\text{EF}} \prod_{{\iqb}}\Big(\xi_{2{\iqb}} a^{\dagger}_{2{\iqb}} +  \xi_{2{\iqb}+1}a^{\dagger}_{2{\iqb}+1}  \Big)\lvert 0 \rangle\\
            &=\prod_{{\iqb}}\Big(\xi_{2{\iqb}} \hat{\text{EF}}  a^{\dagger}_{2{\iqb}}\hat{\text{EF}}^{\dagger} + \xi_{2{\iqb}+1} \hat{\text{EF}} a^{\dagger}_{2{\iqb}+1} \hat{\text{EF}}^{\dagger} \Big)\lvert 0 \rangle\\
            &=\prod_{{\iqb}}\left(\sum_l \Big(\xi_{2{\iqb}} (\eftm)_{2{\iqb},l}  + \xi_{2{\iqb}+1} (\eftm)_{2{\iqb}+1,l}\Big)d^{\dagger}_{l} \right)\lvert 0 \rangle\, .
          \end{split}
          \label{eqn:ef_out}
        \end{equation}
        \par
        In the following subsections, to implement a quantum gate with an 8f-processor~$\hat{\text{EF}}= \hat{\text{FF}}^{(2)} \dop  \hat{\text{FF}}^{(1)}$, we use optimization techniques (see appendix~\ref{sec:opt}) to find appropriate periodic pupil phase function~$\phi_p(x)$ (Eq.~\eqref{eqn:ft[ffp]main}) for the 4f-processors~$ \hat{\text{FF}}^{(1)}$ and~$\hat{\text{FF}}^{(2)}$, and step phase function~$\phi_D(x)$ for the lattice-like spatial modulator~$\dop$, where $\phi_r=\phi_{D}(r l_x)$ (see Eq.~\eqref{eqn:u4fop1d}).
        Furthermore, in the optimizations, we assume that the two 4f-processors~$ \hat{\text{FF}}^{(1)}$ and~$\hat{\text{FF}}^{(2)}$ are equivalent and have similar pupil phase functions~$\phi^{(1)}_p(x)=\phi^{(2)}_p(x)=\phi_p(x)$.
        \subsection{Hadamard Gate}
        \begin{figure*}[tp!]
          \includegraphics[width= \textwidth]{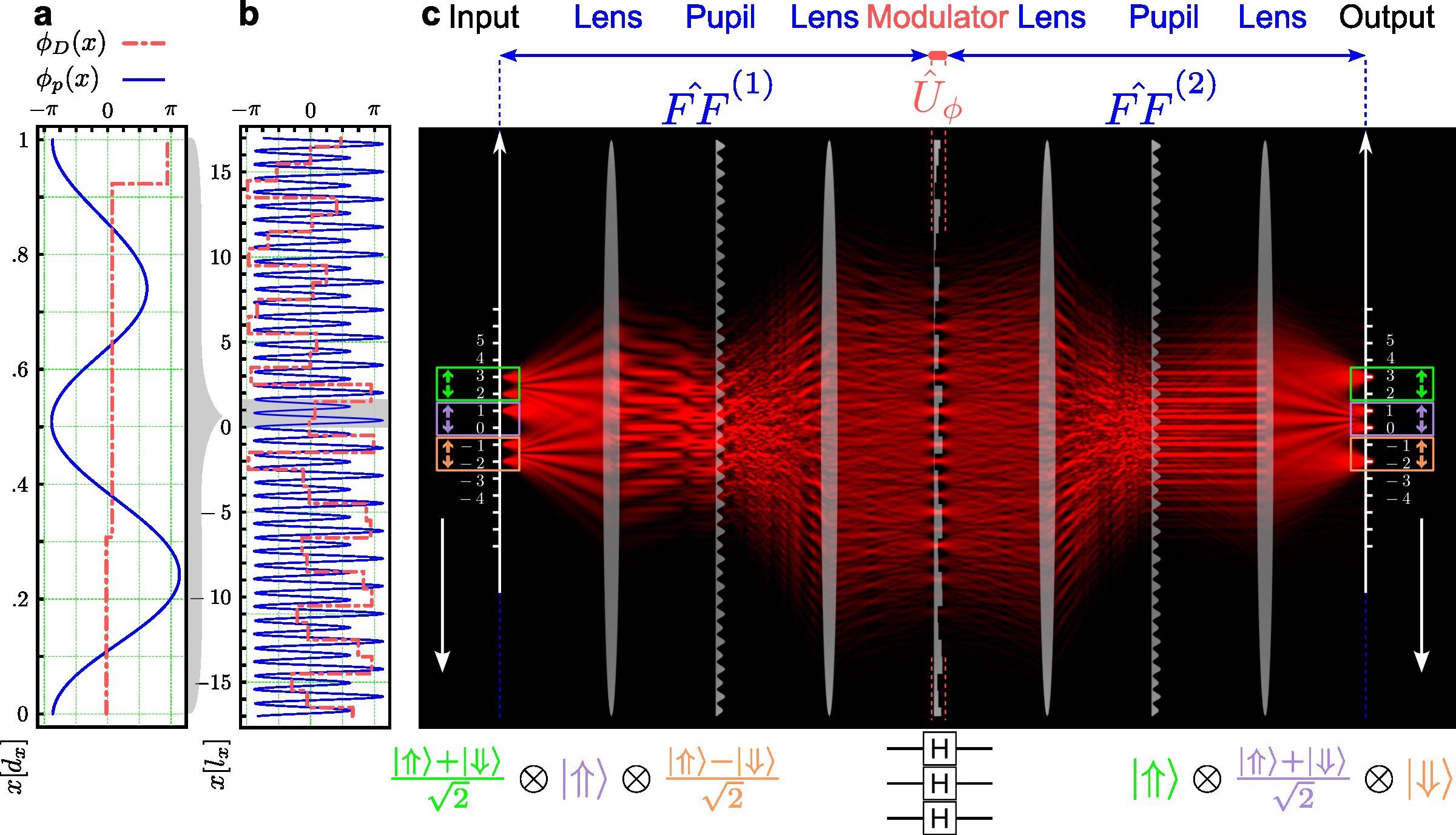}
          \caption{\textbf{Hadamard Gate Implementation via an 8f-Processor.}
          This Hadamard 8f-processor is made up of two similar 4f-processors with periodic pupil phase modulation~$\phi_p (x)$, shown in \textbf{b}. 
          At the interface of these two similar 4f-processors, there is a spatial modulator with step phase function~$\phi_D(x)$, as displayed in \textbf{b}.
          \textbf{a}  zooms out the phase functions of the 4f-systems’ pupils over one period.
          The 8f-processor performs the Hadamard gate on three qubits associated with ports $(3,2), (1,0)$, and~$(-1,-2)$.
          The propagation simulation,  as an example, takes three single photons whit quantum states~$\lvert +\rangle=\frac{1}{\sqrt{2}}\left(\lvert 1 \rangle_{3}+\lvert 1 \rangle_{2}\right)$, $\lvert \Uparrow \rangle=\lvert 1 \rangle_{1}$, and $\lvert -\rangle=\frac{1}{\sqrt{2}}\left(\lvert 1 \rangle_{-1}-\lvert 1 \rangle_{-2}\right)$, as the input of the gate.
          }\label{fig:hadamard}
        \end{figure*}
        Hadamard gate performs the following transformation on a single qubit
         \begin{equation}
          \begin{split}
            \mathbf{\hat H}= \frac{1}{\sqrt{2}}
          \begin{pmatrix}
              1 &1 \\
              1&-1
             \end{pmatrix}\, ,
          \end{split}
          \label{eqn:h}
        \end{equation}
        where, the first and second rows and columns are associated with the qubit's basis states up~$\lvert \Uparrow\rangle$ and down~$\lvert \Downarrow\rangle$, respectively.
        In our formalism, for qubit~${\iqb}$, the down-state~$\lvert\Downarrow\rangle_{\iqb}=\lvert 1\rangle_{2{\iqb}}$ corresponds to a single-photon at lattice point~$m=2{\iqb}$, and the up-state~$\lvert\Uparrow\rangle_{\iqb}=\lvert 1\rangle_{2{\iqb}+1}$ corresponds to a single-photon at lattice point~$m'=2{\iqb}+1$.
        In order for the 8f-system operation Eq.~\eqref{eqn:ef_out} performs the Hadamard gate on the qubit~${\iqb}$, 
        the 8f-transformation matrix Eq.~\eqref{eqn:8ftm}'s elements should be~$ (\eftm)_{2{\iqb}+1,2{\iqb}+1}=(\eftm)_{2{\iqb}+1,2{\iqb}}=(\eftm)_{2{\iqb},2{\iqb}+1}=-(\eftm)_{2{\iqb},2{\iqb}}=\frac{1}{\sqrt{2}}e^{i\theta_{\iqb}}$, where the phase~$\theta_n$ is associated with the degree of freedom available in defining a quantum gate. 
        In other words, these four elements of~$\eftm$ form a $2\times 2$ truncated transformation matrix~$\mathbf{T}$, which corresponds to the gate operator operator~$\mathbf{\hat O_{\iqb}}$ on qubit~${\iqb}$, 
        \begin{equation}
          \begin{split}
            \mathbf{\hat O_{\iqb}}= 
          \begin{pmatrix}
               (\eftm)_{2{\iqb}+1,2{\iqb}+1} & (\eftm)_{2{\iqb}+1,2{\iqb}} \\
               (\eftm)_{2{\iqb},2{\iqb}+1}& (\eftm)_{2{\iqb},2{\iqb}}
             \end{pmatrix}\, ,
          \end{split}
          \label{eqn:on}
        \end{equation}
        which for a perfect Hadamard operation, it, up to a constant phase factor equals the Hadamard gate Eq.~\eqref{eqn:h}, i.e., $\mathbf{\hat O_{\iqb}}=e^{i \theta_{\iqb}}\mathbf{\hat H}$.
        The success probability and the fidelity of the implemented Hadamard gate on qubit~${\iqb}$ are defined by~\cite{uskov_pra_2009}
         \begin{equation}
           \begin{split}
             \mathcal{S}&=\frac{\Tr \big( \mathbf{\hat O_{\iqb}^{\dagger}. \hat O_{\iqb}}\big)}{d}\\
             \mathcal{F}&=\frac{\Big| \Tr \big(\mathbf{\hat O_{\iqb}^{\dagger}. \hat H}\big)\Big|^2}{\Tr \big( \mathbf{\hat O_{\iqb}^{\dagger}. \hat O_{\iqb}} \big) \Tr \big( \mathbf{\hat H^{\dagger}.\hat H} \big) }\, ,
          \end{split}
          \label{eqn:}
        \end{equation}
        where $d$ is the matrix dimension of the implemented gate, which for the Hadamard gate becomes 2.
        \par
        Figure~\ref{fig:hadamard} shows three Hadamard gates implemented via an 8f-processor.
        The Hadamard gates operate on the three qubits associated with $\iqb=1,0,-1$.
        We used the optimization procedure (see appendix~\ref{sec:opt}) to find the 4f-pupil's phase function~$\phi_p(x)$ and the spatial phase modulator's step phase function~$\phi_D(x)$ of the 8f-processors to maximize the gate's fidelity and success probability.
        Figure~\ref{fig:hadamard}a,b shows the optimized phase functions corresponding to near unity fidelity and $99\%$ average success probability of the implemented Hadamard gate on the three qubits.         
        Figure~\ref{fig:hadamard}c shows these three qubits' quantum Fourier optical propagation simulation through the optimized 8f-Hadamard gate.        
        The simulation assumes that the qubits associated with~$\iqb=+1,0 ,-1$ are in the quantum states~$\lvert +\rangle_{1}=\frac{1}{\sqrt{2}}\left(\lvert \Uparrow \rangle_{1}+\lvert \Downarrow \rangle_{1}\right)=\frac{1}{\sqrt{2}}\left(\lvert 1 \rangle_{3}+\lvert 1 \rangle_{2}\right)$, $\lvert \Uparrow \rangle_{0}=\lvert 1 \rangle_{1}$, and $\lvert -\rangle_{-1}=\frac{1}{\sqrt{2}}\left(\lvert \Uparrow \rangle_{-1}-\lvert \Downarrow \rangle_{-1}\right)=\frac{1}{\sqrt{2}}\left(\lvert 1 \rangle_{-1}-\lvert 1 \rangle_{-2}\right)$, respectively.
        At the output of the 8f-Hadamard processor, the qubits are transformed to state~$\lvert \Uparrow\rangle_{1}, \lvert +\rangle_{0}$ and~$\lvert \Downarrow\rangle_{-1}$, respectively.
        \par
        The simulation program gives the Fock state representation of the quantum light at each $z$ point of the propagation.
        Figure~\ref{fig:hadamard}c, to picture the state, shows the photon number (intensity) operator expectation value~$\langle \hat n _x \rangle=\langle \hat a^{\dagger}_x \hat a_x\rangle$ at each propagation step.
        Indeed, the qubit's basis states $\lvert \Uparrow\rangle$ and~$\lvert \Downarrow\rangle$ are evident at the input and output of the 8f-processors.
        Furthermore, the interference intensity patterns give clues to discriminating superposition states~$\lvert + \rangle$ and~$\lvert - \rangle$ from each other.
        For example, the central interference line between the $\lvert \Uparrow\rangle$ and ~$\lvert \Downarrow\rangle$ of a qubit is bright if the qubit is in state~$\lvert + \rangle$, and it is dark for state~ $\lvert - \rangle$ due to the constructive and destructive interference induced by the relative phase between the two ports of each qubit, respectively.
        \par
        This subsection demonstrated the single-qubit Hadamard gate's implementation via an 8f-processor.
        Other single-qubit gates, such as the Pauli gates, can similarly be realized.
       	The 8f-processor implemented single qubit gates' fidelity and success probability becomes unity since a single qubit gate corresponds to a transformation matrix of dimension $N=2$, which can be factorized into $2N-1=3$ circulant and diagonal matrices then is perfectly implementable via an 8f-processor.
	Note that the matrix factorization of an $N \times N$ matrix into $2N-1$ diagonal and circulant matrices are the upper limit to matrix factorization and therefore the upper limit to the required optical modules.
	For example, Fig.~\ref{fig:hadamard} used an 8f-processor to implement the Hadamard gate on 3 qubits which correspond to a transformation matrix with dimension~$N=3*2=6$.
	Furthermore, in the following subsection, we show that an 8f-processor is also sufficient to implement two-qubit gates (a matrix of dimension~$N=4$).
        \subsection{CNOT Gate}                
            \begin{figure*}[tp!]
          \includegraphics[width= \textwidth]{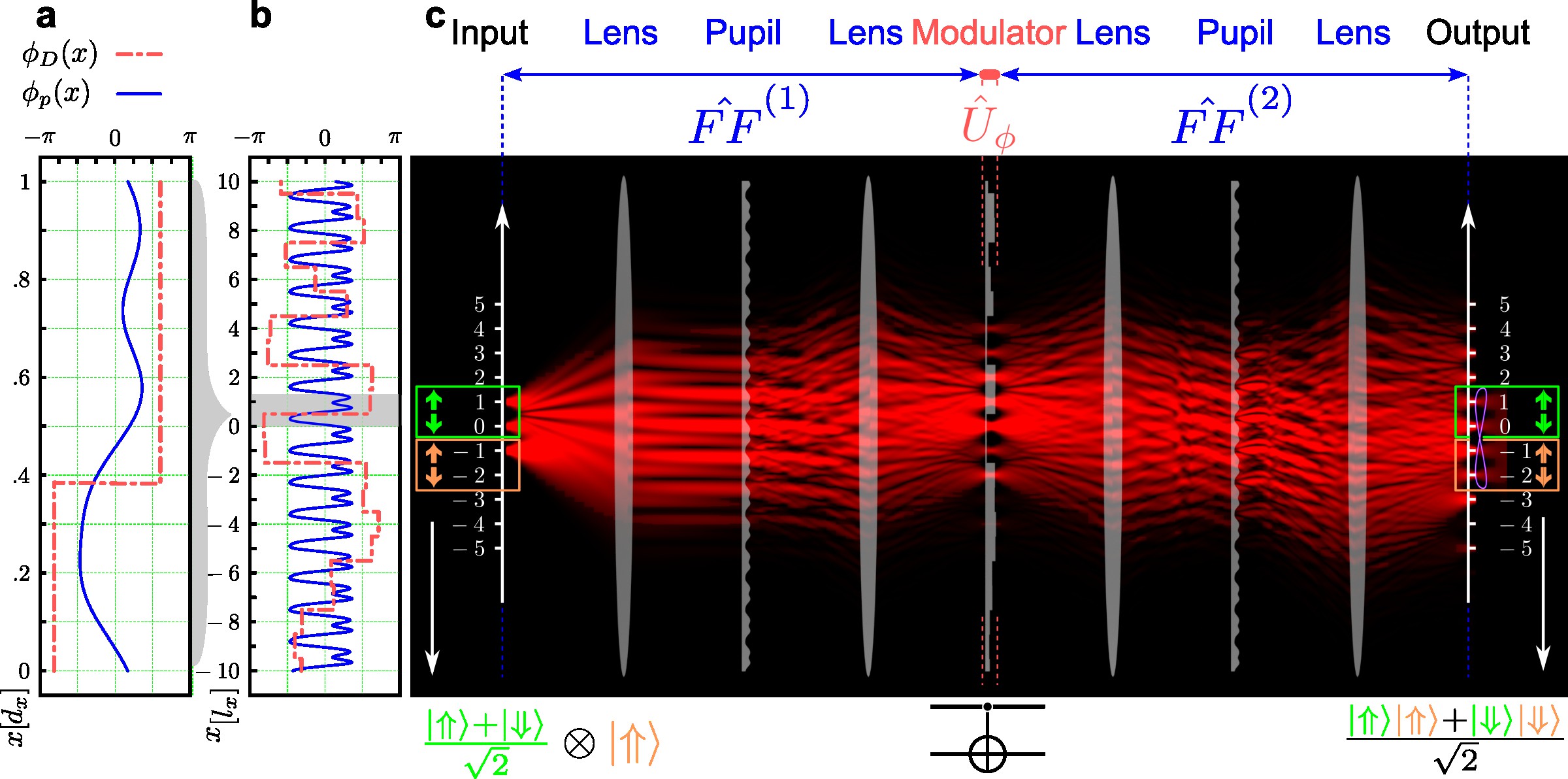}
          \caption{\textbf{CNOT Gate Implementation via an 8f-Processor and a Projective Measurement.}
          Similar to Fig.~\ref{fig:hadamard}, \textbf{a} and \textbf{b} show the gate optimized phase function of the pupils~($\phi_p(x)$) and the modulator($\phi_D(x)$).
          The phases are optimized to implement the CNOT gate on the control qubit associated with ports~$(1,0)$ and the target qubit associated with ports~$(-1,-2)$.
          The propagation simulation takes $\lvert +\rangle=\frac{1}{\sqrt{2}}\left(\lvert 1 \rangle_{1}+\lvert 1 \rangle_{0}\right)$ and ~$\lvert \Uparrow \rangle=\lvert 1 \rangle_{-1}$ as the quantum states of the control and target qubits, respectively.
           The projection operator Eq.~\eqref{eqn:pcnot} is applied to the quantum state of light at the output of the 8f-processor, which entangles the input qubits and reduces the quantum light intensity by a factor of 9.
          }\label{fig:cnot} 
            \end{figure*}
        In addition to the linear optical transformation (e.g., 8f-transformation), which is adequate for implementing single-qubit gates, entangling gates require a projective measurement~$\mathbf{\hat P}$. 
        This projective measurement reduces the output state~$\lvert \Phi \rangle$ (Eq.~\eqref{eqn:ef_out}) of the 8f-processor into state~$\mathbf{\hat P} \lvert \Phi \rangle$.
        Consequently, it reduces the success probability of the gate.
        \par
        In this section, via an 8f-processor and projective measurement, we implement the entangling CNOT gate on the control qubit~${\iqb}$ and the target qubit~${\iqb}'$.
        The matrix representation of the gate is
        \begin{equation}
          \begin{split}
            \mathbf{\widehat {CNOT}}=
          \begin{pmatrix}
            1 &0&0 &0 \\
            0 &1&0 &0 \\
            0 &0&0 &1 \\
            0 &0&1 &0 
             \end{pmatrix}\, ,
          \end{split}
          \label{eqn:cnot}
        \end{equation}
        where columns and rows 1,2,3 and 4 are associated with the two-qubits computational basis states~$\lvert \Uparrow \rangle_{{\iqb}}\lvert \Uparrow \rangle_{{\iqb}'}$,~$\lvert \Uparrow \rangle_{{\iqb}}\lvert \Downarrow \rangle_{{\iqb}'}$,~$\lvert \Downarrow \rangle_{{\iqb}}\lvert \Uparrow \rangle_{{\iqb}'}$~and~$\lvert \Downarrow \rangle_{{\iqb}}\lvert \Downarrow \rangle_{{\iqb}'}$, respectively.
        \par
        To implement the CNOT gate, we use the scheme proposed in~\cite{hofmann_pra_2002,ralph_pra_2002}, which is experimentally favorable to implement~\cite{obrien_n_2003,lu_npj_2019} and, when combined with quantum non-demolition measurements, can be used to implement the KLM protocol~\cite{knill_n_2001}.
        Furthermore, its success probability~$\mathcal{P}=1/9$ is higher than the scheme proposed by Knill~\cite{knill_pra_2002} ($\mathcal{P}=2/27$).
        In the Knill proposal, a demolishing measurement performs on ancilla photons. 
        Therefore, it has the advantage that a direct non-demolition quantum measurement on the computational qubits is not required.
        \par
        In the scheme proposed by~\cite{hofmann_pra_2002,ralph_pra_2002}, the projection is to the quantum light state where only one photon is at the ports associated with the control qubit~${\iqb}$ and one photon at the ports associated with the target qubits~${\iqb}'$, which are ports~$(2{\iqb} ,2{\iqb}+1)$ and $(2{\iqb}',2{\iqb}'+1)$, respectively.
        Therefore, the projection operator is
        \begin{equation}
          \mathbf{\hat P}=\sum\limits_{i=2{\iqb},2{\iqb}+1}\sum\limits_{j=2{\iqb}',2{\iqb}'+1}\tensor[]{\lvert 1\rangle}{_i}\tensor[]{\lvert 1\rangle}{_j} \tensor[_j]{ \langle 1\rvert}{} \tensor[_i]{\langle 1\rvert}{}\, .
          \label{eqn:pcnot}
        \end{equation}
        After applying this projection operator to the output state of the 8f-processor (Eq.~\eqref{eqn:ef_out}), we compose the two-qubit operator~$\mathbf{ \hat O_{{\iqb},{\iqb}'}}$ on qubits~${\iqb}$ and ${\iqb}'$, which is a $4\times4$~matrix with the same basis states' labeling order as the CNOT gate Eq.~\eqref{eqn:cnot}.
        Similar to the Hadamard implementation, we use the optimization process~\cite{uskov_pra_2009} to find the 4f-pupil's phase function~$\phi_p(x)$ and the spatial phase modulator's step phase function~$\phi_D(x)$, which maximize the fidelity and the success probability of the implemented CNOT gate~$\mathbf{\hat O_{{\iqb},{\iqb}'}}$.
        \par
        Figure~\ref{fig:cnot} demonstrates the implementation of a CNOT gate via an 8f-processor on the control qubit~${\iqb}=0$ and the target qubit~${\iqb}'=-1$.
        Figure~\ref{fig:cnot}a,b shows the optimized phase functions~$\phi_p(x)$ and~$\phi_D(x)$ associated with the CNOT gate with fidelity~$\mathcal{F}=0.999$ and success probability~$\mathcal{S}=0.99 \times \frac{1}{9}$ ($\frac{1}{9}$ is the maximum achievable success probability by the above procedure~\cite{hofmann_pra_2002,ralph_pra_2002}).
        Figure~\ref{fig:cnot}c shows the intensity measurement expectation values of the input quantum state propagated through the Fourier optical system.
        The simulation considers quantum states~$\lvert + \rangle_{0}$ and~$\lvert \Uparrow \rangle_{-1}$ for the control and the target qubits, respectively.
        CNOT gate entangles such input qubits. Meaning, that after the projective measurement at the end of the 8f-processor, these two qubits become entangled.
        The simulation shows that the entangled photons coming out of the 8f-processor and the projector does not exhibit interference patterns as before.

\section{Conclusion}
	Optical technology advances led to the development of lattice-like optical fiber arrays. 
	Furthermore, quantum emitter arrays are on the horizon~\cite{berraquero_nc_2017}. 
	These new advancements can make spatially encoded digitalized quantum information sources practically available. 	
	Though, after exiting such sources, quantum light loses its discreteness due to the Huygens–Fresnel principle. 
	However, to implement discrete unitary matrix operation on such lattice-like (discretized, digitalized) wavefronts, the paper introduces a class of 4f-processors that retain the lattice-like structure at their output surface. 
	In other words, the 4f-processors with a periodic phase-only pupil with spatial angular frequencies~$\kappa_x= \frac{\mko \lx}{f}$ and $\kappa_y=\frac{2 \pi }{\yg}= \frac{\mko \ly}{f}$ preserve the input lattice constants~$\lx$ and $\ly$ in the $x$ and~$y$ directions, respectively. 
	Such 4f-processors perform unitary circulant matrix (tensor) operations on the discretized input wavefront. 
	Therefore, Fourier optics allows us to implement universal multiport operations and interferences, given that any matrix can be factorized into a sequence of alternating diagonal and circulant matrices.
	\par
	We use the quantum Fourier optics theory to study the evolution of input quantum states of light composed of various qubits, meaning single-photons with discretized arbitrary wavefronts. 
	Since the introduced quantum Fourier optical structure can perform any multiport unitary operation, it offers a powerful platform for linear quantum computations, such as KLM protocol and Boson sampling.
	 As a demonstration, we implement the single-qubit Hadamard and the two-qubit entangling CNOT gates based on the proposed structures.
	 Furthermore, we also present the simulation of quantum light intensity through the introduced implemented gates.  
	\par
	There are several reasons why Fourier optical quantum computation is a promising candidate for linear optical quantum computation, including the following. 
	The required modules are quadratically fewer than the common implementation approach by beam-splitters and phase-shifters. 
	Compared to the other dimensions of photons, such as path, polarization, frequency, and time, the photon wavefront has a higher capacity for quantum information encoding. 
	The Fourier optical quantum gate can be electrically programmed with programmable spatial light modulators. 
	Thus, the same setup can be used for various quantum algorithms. 
	Programmable Fourier optical setups are more stable over time than programmable Mach-Zehnder implementations.
	There are also many applications of universal multiport transformations in classical domains, such as optical switching and routing, where the proposed scheme might be useful.
         \appendix
        \section{Optimization and Simulation}
        \label{sec:opt}
        This paper assumes phase-only pupils for the 4f-systems.
        Nevertheless, not every circulant matrix representation of a 4f-system corresponds to a phase-only pupil.
        However, since circulant matrices can be written as~$F^{\dagger}.D.F$, where $F$ is the unitary DFT matrix and $D$ is a diagonal matrix, the corresponding diagonal matrix of a unitary circulant matrix is unitary.
        Therefore, a unitary circulant matrix is phase-only at the sampling points.
        \par
        Furthermore, the discrete Fourier transform (DFT) matrix multiplication is computationally favorable.
        Therefore, we decompose a circulant matrix~$\fftm $ as $\fftm =\mathbf{F. D_{FF}.F^{\dagger}}$, where~$\mathbf{F}$ is the DFT matrix, and $\mathbf{D_{FF}}$ is the corresponding diagonal matrix of circulant matrix~$\fftm $.
        The $i$th diagonal element of the matrix, $(\mathbf{D_{FF}})_{ii}$, corresponds to the $i$th sample of the 4f-operator's pupil.
        According to the Nyquist–Shannon sampling theorem, the sampling period of~$\frac{\pi}{\kappa_{max}}=\frac{\pi}{R\kappa_x}$, where $\kappa_{max}=R\kappa_x$ denotes the maximum spatial angular frequency of the pupil, is enough to avoid aliasing.
        Since the pupil is periodic with period~$d_x=\frac{2\pi}{\kappa_x}$, $2R$ samples are theoretically sufficient to find the pupil's phase factor~$P(x)=e^{-i \phi_p(x)}$.
        \par
      We use the numerical optimization procedure~\cite{uskov_pra_2009} to find the required phase for the desired quantum gate.
      In this optimization approach, we first maximize the fidelity~$\mathcal{F}$ of the implemented gate to near unity. In the second step, the gate's success probability~$\mathcal{S}$ is optimized by maximizing the penalty function~$\mathcal{F}+\mu \mathcal{S}$ on the manifold of the unity fidelity.
      We modify the optimization parameter~$\mu$ to get the best result.  
      \par
      If the phase functions~$\phi_p^{(i)}(x)$ of 4f-operators~$i=1,2$ are periodic, their corresponding phase factors~$P^{(i)}(x)=e^{-i \phi_p^{(i)}(x)}$ become periodic with the same period.
      Therefore, we consider the Fourier coefficients of the phase functions~$\phi_{p}^{(i)}(x)$ as the 4f-operator's variables  to be optimized, which are the coefficients~$S^{(i)}_n$ and $C^{(i)}_n$ of the Fourier expansion of the pupils' phase functions~$\phi^{(i)}_p(x)=\sum_{n} \left( S^{(i)}_n\sin{n \kappa_x x}+C^{(i)}_n\cos{n \kappa_x x} \right)$.
       The amplitudes of the spatial phase modulator's step phase function~$\phi_D(x)$, i.e.,  $\phi_r$ in Eq.~\eqref{eqn:ubud}, are also the optimization variables.
      \par
	For the Hadamard matrix, the optimized phase functions are shown in Fig.~\ref{fig:hadamard} a,b.
	From this phase functions and Eq.~\eqref{eqn:efanef}-\eqref{eqn:8ftm}, one can find the 8f-transformation matrix on the field operators.
	The  truncation of this transformation matrix to elements~$n,l=2{\iqb},2{\iqb}+1$ gives the transformation matrix~$\mathbf{T_{{\iqb}}}$ associated with qubit~${\iqb}$. 
	The single-qubit gate operator~$\mathbf{\hat O_{\iqb}}$ equals the transformation matrix~$\mathbf{T_{{\iqb}}}$.
	This optimization procedure gives the following gate operators~$\mathbf{\hat O_{\iqb}}$ for the qubits~${\iqb}=+1,0,-1$
	\ifdefined \showhelp
	$iis= [3 2] jjs= [3 2] \qquad  spH= 0.991241 fidH= 0.9999907412558193$
	\newline
	$ iis= [1 0] jjs= [1 0] \qquad  spH= 0.989829 fidH= 0.9999940914954335$
	\newline
	 $iis= [-1 -2] jjs= [-1 -2]\qquad spH= 0.989838 fidH= 0.9999863302487604$	
	\fi
\begin{equation} 
          \begin{split}
            \mathbf{\hat O_{+1}}&=
          \begin{pmatrix}
			0.705 e^{0.131i} & 0.702 e^{0.13i}  \\
 			0.702 e^{0.13i} & 0.707 e^{-3.011i}  
			\end{pmatrix} \, , \\
            \mathbf{\hat O_{0}}&=
          \begin{pmatrix}
			0.705 e^{-3.085i} & 0.702 e^{-3.083i}  \\
			 0.702 e^{-3.083i} & 0.705 e^{0.056i} 
			\end{pmatrix} \, , \\
            \mathbf{\hat O_{-1}}&=
          \begin{pmatrix}
			0.707 e^{-0.035i} & 0.701 e^{-0.035i} \\
 			 0.701 e^{-0.035i} & 0.705 e^{3.107i} 
			\end{pmatrix}\, .
          \end{split}
          \label{eqn:}
        \end{equation}
        These optimized Hadamard Gates have the fidelity $\mathcal{F}=0.99999$ and the average success probability of $\mathcal{S}=0.99$.
        \par
        Figure~\ref{fig:cnot}a,b shows the optimized phases of the 8f-processor, which applies the CNOT gate on the control qubit~${\iqb}=0$ and target qubit~${\iqb}'=-1$.
        The filed operators for these qubits are $\hat a^{\dagger}_{+1},\hat a^{\dagger}_{0}$ and $\hat a^{\dagger}_{-1},\hat a^{\dagger}_{-2}$, respectively. 
        The optimized  truncated transformation matrix on them is as follows	
	  \small
	\begin{equation}
          \begin{split}
            \textbf{T}=
          	\begin{pmatrix}
	 		0.58 e^{2.758i} & 0.0 e^{-0.642i} & 0.004 e^{2.993i} & 0.005 e^{-2.713i} \\
			0.0 e^{-0.642i} & 0.568 e^{2.762i} & 0.576 e^{-1.25i} & 0.57 e^{1.89i} \\
 			0.004 e^{2.993i} & 0.576 e^{-1.25i} & 0.579 e^{-2.115i} &0.006 e^{-2.506i} \\
 			0.005 e^{-2.713i} & 0.57 e^{1.89i} & 0.006 e^{-2.506i} &''0.574 e^{-2.125i} 
 		\end{pmatrix}\, .
          \end{split}
          \label{eqn:tcnot}
\end{equation}
\normalsize
	See Eq.~\eqref{eqn:heispicffa} for the arrangement and representation of the field operators' vectors and transformation matrices used in this paper, which is from the higher to lower indices.
        The optimized 8f-transformation matrix~Eq.~\eqref{eqn:tcnot}, together with the projective measurement~\eqref{eqn:pcnot}, gives the optimized CNOT gate operator~$\mathbf{ \hat O_{{\iqb},{\iqb}'}}$  on qubits~${\iqb}=0$ and ${\iqb}'=-1$ as follows
 \small
  \begin{equation}
          \begin{split}
            \mathbf{ \hat O_{{0},{-1}}}=
          \begin{pmatrix}
		0.336 e^{0.643i} & 0.003 e^{0.252i} & 0.002 e^{1.743i} &0.003 e^{2.32i} \\
 		0.003 e^{0.252i} & 0.333 e^{0.633i} & 0.002 e^{-1.4i} &0.003 e^{-0.823i} \\
		0.002 e^{1.743i} & 0.002 e^{-1.4i} & 0.003 e^{-3.049i} & 0.331 e^{0.636i}  \\
		0.003 e^{2.32i} & 0.003 e^{-0.823i} & 0.331 e^{0.636i} & 0.001 e^{0.253i}  
 	\end{pmatrix}\, ,
          \end{split}
          \label{eqn:}
	\end{equation}
	\normalsize
	which gives fidelity of~$\mathcal{F}=0.999$ and success probability of $\mathcal{S}=0.99*\frac{1}{9}$.
       \par
      We use our Python scripts to run the simulation (see also~\cite{qfo}).
      It considers each state of the qubits as a Gaussian beam (Gaussian photon-wavepacket)~\cite{kok_cambridge_2010} with a width of~$\approx \SI{10}{\micro\metre} $.
      The other parameters of the simulation are as follows: the lattice distance~$l_x=\SI{100}{\micro\metre}$, wavelength $\lambda= \SI{650}{\nano\metre} $, the lenses' focal length~$f=\SI{2.5}{\centi\metre}$ for Fig.~\ref{fig:hadamard} and~$f=\SI{2}{\centi\metre} $ for Fig.~\ref{fig:cnot}.
      These parameters satisfy the paraxial approximation~$(k_x/k)^2 < 0.01$.
      The Python script creates an OpenVDB file from the propagation data.
      The file is imported into the Blender software and visualized.

\end{document}